\documentclass[aps,twocolumn,superscriptaddress,prl,bibnotes]{revtex4}

\usepackage{graphicx}
\usepackage{revsymb}
\begin{document}


\title{Valley splitting in Si quantum dots embedded in SiGe}

\author{S. Srinivasan}
\author{G. Klimeck}
\affiliation{Network for Computational Nanotechnology, Purdue University,
West Lafayette, Indiana
47907 USA}
\affiliation{Birck Nanotechnology Center, Purdue University,
West Lafayette, Indiana
47907 USA}
\author{L.~P. Rokhinson}
\affiliation{Birck Nanotechnology Center, Purdue University, West Lafayette,
Indiana 47907 USA}
\affiliation{Department of Physics, Purdue University, West
Lafayette, Indiana 47907 USA}

\begin{abstract}
We examine energy spectra of Si quantum dots embedded into
Si$_{0.75}$Ge$_{0.25}$ buffers using atomistic numerical calculations for
dimensions relevant to qubit implementations. The valley degeneracy of the
lowest orbital state is lifted and valley splitting fluctuates with monolayer
frequency as a function of the dot thickness. For dot thicknesses $\leq6$ nm
valley splitting is found to be $>150\ \mu$eV. Using the unique advantage of
atomistic calculations we analyze the effect of buffer disorder on valley
splitting. Disorder in the buffer leads to the suppression of valley splitting
by a factor of 2.5, the splitting fluctuates with $\approx20\mu$eV for
different disorder realizations.  Through these simulations we can guide future
experiments into regions of low device-to-device fluctuations.
\end{abstract}

\pacs{}

\maketitle

Understanding and design of silicon nanometer-scaled electronic devices has
regained significant interest.  This interest is sparked by the experimental
progress that enabled the reproducible construction of geometries in which
electrons are confined in three dimensions to length scales of a few nanometers
and the potential applications of this technology to ultra-scaled traditional
CMOS devices. Emerging application of Si nanostructures for qubit
implementations due to long spin relaxation
times\cite{kane98,wilamowski02,tyryshkin05} imposes additional stringent
requirements on energy spectrum engineering, including the precise control of
valley degeneracy.  The six-fold valley degeneracy of bulk Si is reduced to
two-fold degeneracy when electrons are confined to two dimensions (2D), such as
at Si/SiO$_2$ interface in mainstream MOSFETs. Already decades ago it was
recognized that there is a small splitting between the two valleys in the
lowest subband\cite{ando82}. Recently, calculations predicted that valley
splitting in narrow (few nm) SiGe/Si/SiGe quantum wells can be of the order of
10-100 meV and should fluctuate rapidly with the well
thickness\cite{boykin04,boykin04a,nestoklon06,friesen07}. However,
experiments\cite{weitz96,koester97,lai04} produced valley splitting about 2
orders of magnitude smaller than that prediction, which has been
explained\cite{kharche07} by the disorders of the Si/SiGe interface and in the
SiGe buffer. The experiments\cite{goswami07} and theoretical methods indicated
that additional spatial confinement will minimize the role of interface
disorder and increase valley splitting. In this paper we investigate the role
of SiGe buffer disorder on valley splitting and answer the fundamental question
of the size and controllability of valley splitting for relevant experimental
structures.

Three dimensional (3D) confinement of electrons can be achieved by various
techniques. Electrostatic surface gating of 2D gas provides relatively weak and
smooth spatial confinement potentials. In contrast, 3D confinement by
Si/SiO$_2$ interface produces sharp potential with Coulomb energies approaching
room temperature\cite{takahashi95,zhuang98} and large valley
splitting\cite{rokhinson02}. Recently, an alternative approach to 3D
confinement has been demonstrated with an advantage of lithographically defined
epitaxial Si/SiGe interfaces using post-fabrication regrowth\cite{bo06}. In
this case spurious charging effects\cite{rokhinson00a}, related to the traps in
SiO$_2$ or unpassivated interface can be avoided, yet retaining sharp confining
potential. We will simulate such defined Si nanostructures in SiGe buffers and
explore sizes relevant for qubit implementations. Simulation capabilities to
represent structures containing 10 million atoms explicitly enable the atomic
representation of the dot, interfaces and the SiGe buffer. Atomistic
simulations also present a unique opportunity to vary the amount of the buffer
disorder in order to attain detailed understanding of the physics of valley
splitting, including its magnitude and fluctuations. The valley splitting is
primarily defined by the smallest dimension of the device and our conclusions
are applicable to any Si nanostructure defined from SiGe/Si/SiGe quantum wells.

\begin{figure}[t]
\def\ffile{unitcell}
\includegraphics[scale=0.3]{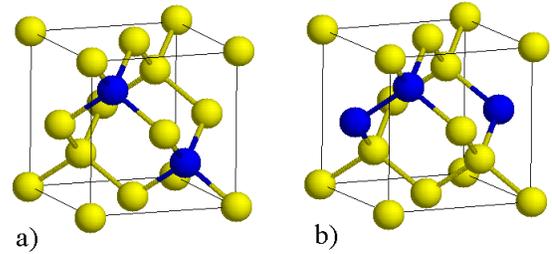}
\caption{Placement of Si (yellow) and Ge (blue) atoms in a) fully ordered
(Si-Ge and
Si-Si bonds) and b) partially ordered (Si-Ge, Ge-Ge and Si-Si bonds) 8-atom
supercells.}
\label{\ffile}
\end{figure}

Calculations of the energy spectrum are performed using the NEMO-3D general
purpose code, which represents each atom in the domain explicitly. The theory
undelying the tool and its relevant benchmarks are given in
Refs~\cite{klimeck02,klimeck07}. The structure is defined on the relaxed (001)
Si$_{0.75}$Ge$_{0.25}$ substrate and the Keating valence-force field model is
used to adjust atomic positions to minimize the strain energy. Calculations of
electronic structure are based on the 20 band $sp^3d^5s^*$ tight-binding model.
The quantum dot was modeled as a $l_x\times l_y\times l_z$ rectangle grown on
37 nm-thick substrate and embedded into 27 nm-thick Si$_{0.75}$Ge$_{0.25}$
buffer, $l_z <l_x,l_y$, where $z$ is along the growth direction. We
investigated the influence of the buffer thickness on electronic structure,
there were no significant changes for substrates $t_s>30$ nm and buffers
$t_b>20$ nm.

For 25\% Ge we can generate various placements of Ge atoms in the
Si$_{0.75}$Ge$_{0.25}$ buffer, with fully ordered containing only Si-Ge bonds,
partially ordered containing single Ge-Ge bond per 8-atom supercell in a fixed
position, and disordered having random placement of Ge atoms retaining 25\%
composition, see schematic in Fig.~\ref{unitcell}.

\begin{figure}[t]
\def\ffile{levels}
\includegraphics[scale=.55]{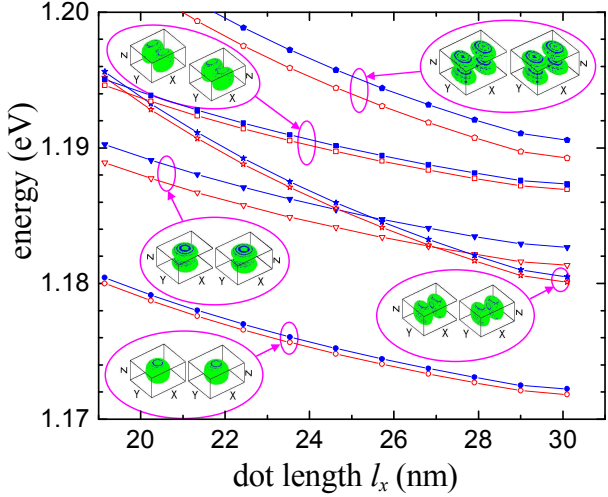}
\caption{Energy levels in $l_x\times20$ nm $\times10$ nm Si dot embedded into
ordered
Si$_{0.75}$Ge$_{0.25}$ buffer.  Energies are referenced to the valence band
$\Gamma^8_v$
point. Inserts show spatial distribution of wavefunctions for the lowest
levels.}
\label{\ffile}
\end{figure}

We start with the analysis of energy levels and valley splitting in a dot
embedded into a fully ordered buffer. Evolution of energy levels for a
$l_x\times20$ nm$ \times 10$ nm dot is shown in Fig.~\ref{levels} (the actual
dot thickness $l_z=9.85$nm $=72$ monolayers). All levels come in pairs, both
levels in the pair having similar wavefunction envelopes (each level is also
double spin-degenerate, which has been confirmed by calculations and will be
ignored for the rest of the paper). The 3D representations of the envelope
wavefunctions at 20\% value are shown for the lowest 6 levels. The two lowest
levels have similar $s$-type wavefunctions and represent the same orbital state
with different valley number. The energy difference between them we call valley
splitting $\Delta^0_v$. The next two levels have one node and belong to the
next orbital state. For $l_x<25$ nm the $p_z$-type state has lower energy than
$p_x$- and $p_y$-type states due to the combination of sizes and effective mass
anisotropy. The $p_x$-type level has the highest sensitivity to $l_x$, as
expected, and for $l_x>26$ nm its energy becomes lower than that of the
$p_z$-type state. Energy separation between the ground and the first exited
orbital states $\delta E\approx8-10$ meV is large enough to restrict qubit
Hilbert space to the lowest orbital state at low temperatures.

Valley mixing results from superposition of two counter-propagating waves
reflected from the opposite Si/SiGe heterointerfaces of the dot. The phase
difference of the two waves depends on the details of the interface. The
strength of the mixing depends on the amplitude of the wavefunctions at the
interfaces, $\Delta_v\propto |\chi(l_b)|^2$, where $\chi(l_b)$ is the value of
the envelope of the electron wavefunction at the dot
boundary\cite{nestoklon06}. For $p_z$-type and $d_z$-type (top curve in
Fig.~\ref{levels}) states wavefunctions are pushed toward $z$-heterointerface
and valley splitting for these state are significantly larger than for the
ground and $p_x$- or $p_y$-type states.

\begin{figure}[t]
\def\ffile{valleyx}
\hspace{-0.2in}
\includegraphics[scale=0.67]{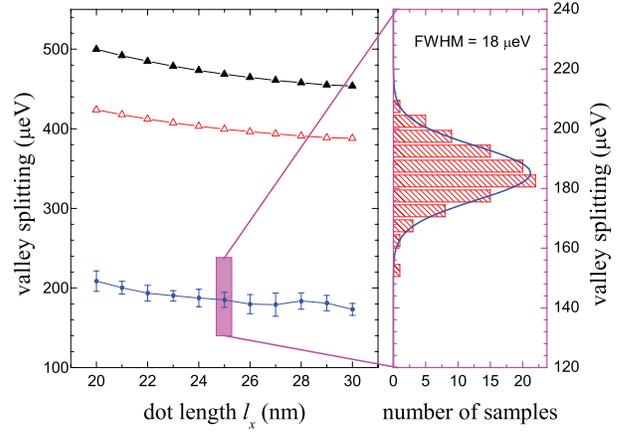}
\caption{Valley splitting for the lowest orbital level as a function of the
dot
size for
ordered (black), partially ordered (red) and disordered (blue)
Si$_{0.75}$Ge$_{0.25}$
buffer. Bars indicate std deviation for each point. An example of valley
splitting
distribution for 100 realizations of buffer disorder is shown in the histogram
for
$l_x=25$ nm, blue curve is the Gaussian fit.}
\label{\ffile}
\end{figure}

The most interesting question which can be uniquely studied by atomistic
calculations is the role of buffer disorder. In Fig.~\ref{valleyx} valley
splitting of the ground level is plotted for a $l_x\times20$ nm$\times10$ nm
dot as a function of the dot size $l_x$ for fully ordered, partially ordered
and completely disordered buffer. For fully ordered buffer the valley splitting
is $\sim0.5$ meV, consistent with analytical calculations. The value does not
change significantly with the dot size, which confirms that valley splitting is
primarily determined by the smallest dimension. For partially ordered buffers
we see a reduction of $\Delta^0_v$ by 10\%, while for fully disordered buffer
$\Delta^0_v$ is reduced 2.5 times to $\sim0.2$ meV. To investigate fundamental
reproducibility of $\Delta^0_v$ we performed calculations for 100 realizations
of the buffer disorder for each point. The histogram of $\Delta^0_v$ for
$l_x=25$ nm dot is plotted in the right frame. The distribution is Gaussian,
with standard deviation of 9.4 $\mu$eV, which is $\sim5$\% of $\Delta^0_v$. The
bars on the main plot indicate standard deviation for other dot sizes.

\begin{figure}[t]
\def\ffile{valleyz}
\includegraphics[scale=0.85]{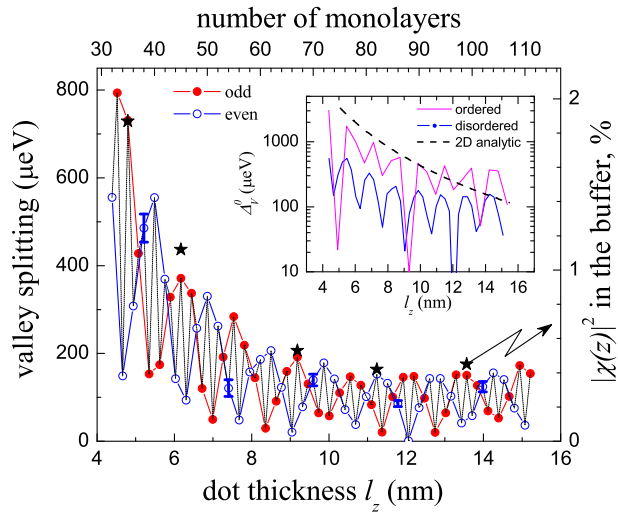}
\caption{Valley splitting for the lowest orbital level of 25 nm $\times\ 20$
nm
$\times\
l_z$ nm Si dot as a function of the dot thickness in monolayers (ML). $l_z$ is
calculated using 1ML$\approx0.13707$ nm. Thin line connects points 1 ML apart, thick lines
connect points 2 MLs apart (open dots for even MLs and solid dots for odd MLs).
Bars indicate std deviations for different disorder realizations. Stars show
percent of the wavefunction
penetrating into the buffer in $z$ direction.
In the inset $\Delta^0_v$ for ordered and disordered buffers are plotted.
Dashed line is $\Delta^0_v$ obtained analytically
for the 2D case.}
\label{\ffile}
\end{figure}

Inter-valley mixing is very sensitive to the smallest dimension of the dot,
$l_z$, and fluctuates with a monolayer (ML) frequency $\Delta_v\propto
\cos(k_0l_z)$, where $k_0=0.82(2\pi/a)$ is the center of the valleys and $a$ is
the lattice constant. Valley splitting as a function of $l_z$ with ML
resolution is plotted in Fig.~\ref{valleyz} (black line), and bars indicate
standard deviation for different disorder realizations. It has been noted that
QWs with odd and even number of MLs belong to different symmetry
classes\cite{nestoklon06}. Indeed, if we connect $\Delta_v^0$ for even and odd
number of MLs we obtain two similar curves which fluctuate with a period of
$\approx8$ MLs and are out-of-phase with each other. The value of
$\Delta_v^0(l_z)$ for the dot embedded into a disordered buffer is reduced by a
factor of 2.5, as shown in the inset. For comparison we also plot valley
splitting calculated for the 2D QW using envelope function
method\cite{nestoklon06} (dashed line), which coincides with our calculations
for the ordered buffer. Saturation of valley splitting for large $l_z$,
compared with the $1/l_z^3$ analytical dependence, is due to an additional
lateral confinement. In Fig.~\ref{valleyz} stars indicate percentage of the
wavefunction $|\chi(z)|^2$ which penetrates the buffer above and below the dot,
the envelope of $\Delta^0_v$ follows $|\chi(z)|^2$ as a function of $l_z$.

To summarize, we calculate energy levels and valley splitting for a small Si
dot embedded in a disordered Si$_0.75$Ge$_0.25$ buffer. We find that buffer
disorder leads to the suppression of valley splitting by $\sim 2.5$ and actual
values fluctuate with  standard deviation of $\sim 20\ \mu$eV. At the same time
disorder limits the lowest valley splitting, which can reach zero for a
perfectly ordered buffer for some dot thicknesses, and dots with valley
splitting $>150\ \mu$eV can be predictably designed from narrow QW ($l_z\leq6$
nm).

The work was supported by ARO/LPS Award No. W911NF-05-1-0437. The use of
nanoHUB.org computational resources operated by the Network for Computational
nanotechnology funded by NSF is acknowledged.

\bibliography{SiGeQD,rohi}

\begin{thebibliography}{20}
\expandafter\ifx\csname natexlab\endcsname\relax\def\natexlab#1{#1}\fi
\expandafter\ifx\csname bibnamefont\endcsname\relax
  \def\bibnamefont#1{#1}\fi
\expandafter\ifx\csname bibfnamefont\endcsname\relax
  \def\bibfnamefont#1{#1}\fi
\expandafter\ifx\csname citenamefont\endcsname\relax
  \def\citenamefont#1{#1}\fi
\expandafter\ifx\csname url\endcsname\relax
  \def\url#1{\texttt{#1}}\fi
\expandafter\ifx\csname urlprefix\endcsname\relax\def\urlprefix{URL }\fi
\providecommand{\bibinfo}[2]{#2}
\providecommand{\eprint}[2][]{\url{#2}}

\bibitem[{\citenamefont{Kane}(1998)}]{kane98}
\bibinfo{author}{\bibfnamefont{B.}~\bibnamefont{Kane}},
  \bibinfo{journal}{Nature (London)} \textbf{\bibinfo{volume}{393}},
  \bibinfo{pages}{133} (\bibinfo{year}{1998}).

\bibitem[{\citenamefont{Wilamowski et~al.}(2002)\citenamefont{Wilamowski,
  Jantsch, Malissa, and R{\"o}ssler}}]{wilamowski02}
\bibinfo{author}{\bibfnamefont{Z.}~\bibnamefont{Wilamowski}},
  \bibinfo{author}{\bibfnamefont{W.}~\bibnamefont{Jantsch}},
  \bibinfo{author}{\bibfnamefont{H.}~\bibnamefont{Malissa}}, \bibnamefont{and}
  \bibinfo{author}{\bibfnamefont{U.}~\bibnamefont{R{\"o}ssler}},
  \bibinfo{journal}{\prb} \textbf{\bibinfo{volume}{66}},
  \bibinfo{pages}{195315} (\bibinfo{year}{2002}).

\bibitem[{\citenamefont{Tyryshkin et~al.}(2005)\citenamefont{Tyryshkin, Lyon,
  Jantsch, and Sch{\"a}ffler}}]{tyryshkin05}
\bibinfo{author}{\bibfnamefont{A.}~\bibnamefont{Tyryshkin}},
  \bibinfo{author}{\bibfnamefont{S.}~\bibnamefont{Lyon}},
  \bibinfo{author}{\bibfnamefont{W.}~\bibnamefont{Jantsch}}, \bibnamefont{and}
  \bibinfo{author}{\bibfnamefont{F.}~\bibnamefont{Sch{\"a}ffler}},
  \bibinfo{journal}{\prl} \textbf{\bibinfo{volume}{94}},
  \bibinfo{pages}{126802} (\bibinfo{year}{2005}).

\bibitem[{\citenamefont{Ando et~al.}(1982)\citenamefont{Ando, Fowler, and
  Stern}}]{ando82}
\bibinfo{author}{\bibfnamefont{T.}~\bibnamefont{Ando}},
  \bibinfo{author}{\bibfnamefont{A.~B.} \bibnamefont{Fowler}},
  \bibnamefont{and} \bibinfo{author}{\bibfnamefont{F.}~\bibnamefont{Stern}},
  \bibinfo{journal}{Rev. of Mod. Phys.} \textbf{\bibinfo{volume}{54}},
  \bibinfo{pages}{437} (\bibinfo{year}{1982}).

\bibitem[{\citenamefont{Boykin et~al.}(2004{\natexlab{a}})\citenamefont{Boykin,
  Klimeck, Eriksson, Friesen, Coppersmith, von Allmen, Oyafuso, and
  Lee}}]{boykin04}
\bibinfo{author}{\bibfnamefont{T.~B.} \bibnamefont{Boykin}},
  \bibinfo{author}{\bibfnamefont{G.}~\bibnamefont{Klimeck}},
  \bibinfo{author}{\bibfnamefont{M.~A.} \bibnamefont{Eriksson}},
  \bibinfo{author}{\bibfnamefont{M.}~\bibnamefont{Friesen}},
  \bibinfo{author}{\bibfnamefont{S.~N.} \bibnamefont{Coppersmith}},
  \bibinfo{author}{\bibfnamefont{P.}~\bibnamefont{von Allmen}},
  \bibinfo{author}{\bibfnamefont{F.}~\bibnamefont{Oyafuso}}, \bibnamefont{and}
  \bibinfo{author}{\bibfnamefont{S.}~\bibnamefont{Lee}},
  \bibinfo{journal}{Appl. Phys. Lett.} \textbf{\bibinfo{volume}{84}},
  \bibinfo{pages}{115} (\bibinfo{year}{2004}{\natexlab{a}}).

\bibitem[{\citenamefont{Boykin et~al.}(2004{\natexlab{b}})\citenamefont{Boykin,
  Klimeck, Friesen, Coppersmith, von Allmen, Oyafuso, and Lee}}]{boykin04a}
\bibinfo{author}{\bibfnamefont{T.~B.} \bibnamefont{Boykin}},
  \bibinfo{author}{\bibfnamefont{G.}~\bibnamefont{Klimeck}},
  \bibinfo{author}{\bibfnamefont{M.}~\bibnamefont{Friesen}},
  \bibinfo{author}{\bibfnamefont{S.~N.} \bibnamefont{Coppersmith}},
  \bibinfo{author}{\bibfnamefont{P.}~\bibnamefont{von Allmen}},
  \bibinfo{author}{\bibfnamefont{F.}~\bibnamefont{Oyafuso}}, \bibnamefont{and}
  \bibinfo{author}{\bibfnamefont{S.}~\bibnamefont{Lee}},
  \bibinfo{journal}{Phys. Rev. B} \textbf{\bibinfo{volume}{70}},
  \bibinfo{pages}{165325} (\bibinfo{year}{2004}{\natexlab{b}}).

\bibitem[{\citenamefont{Nestoklon et~al.}(2006)\citenamefont{Nestoklon, Golub,
  and Ivchenko}}]{nestoklon06}
\bibinfo{author}{\bibfnamefont{M.~O.} \bibnamefont{Nestoklon}},
  \bibinfo{author}{\bibfnamefont{L.~E.} \bibnamefont{Golub}}, \bibnamefont{and}
  \bibinfo{author}{\bibfnamefont{E.~L.} \bibnamefont{Ivchenko}},
  \bibinfo{journal}{Phys. Rev. B} \textbf{\bibinfo{volume}{73}},
  \bibinfo{pages}{235334} (\bibinfo{year}{2006}).

\bibitem[{\citenamefont{Friesen et~al.}(2007)\citenamefont{Friesen, Chutia,
  Tahan, and Coppersmith}}]{friesen07}
\bibinfo{author}{\bibfnamefont{M.}~\bibnamefont{Friesen}},
  \bibinfo{author}{\bibfnamefont{S.}~\bibnamefont{Chutia}},
  \bibinfo{author}{\bibfnamefont{C.}~\bibnamefont{Tahan}}, \bibnamefont{and}
  \bibinfo{author}{\bibfnamefont{S.~N.} \bibnamefont{Coppersmith}},
  \bibinfo{journal}{Phys. Rev. B} \textbf{\bibinfo{volume}{75}},
  \bibinfo{pages}{115318} (\bibinfo{year}{2007}).

\bibitem[{\citenamefont{Weitz et~al.}(1996)\citenamefont{Weitz, Hauga,
  Klitzing, and Scha\"ffler}}]{weitz96}
\bibinfo{author}{\bibfnamefont{P.}~\bibnamefont{Weitz}},
  \bibinfo{author}{\bibfnamefont{R.}~\bibnamefont{Hauga}},
  \bibinfo{author}{\bibfnamefont{K.~V.} \bibnamefont{Klitzing}},
  \bibnamefont{and}
  \bibinfo{author}{\bibfnamefont{F.}~\bibnamefont{Scha\"ffler}},
  \bibinfo{journal}{Surf. Sci.} \textbf{\bibinfo{volume}{361/362}},
  \bibinfo{pages}{542} (\bibinfo{year}{1996}).

\bibitem[{\citenamefont{Koester et~al.}(1997)\citenamefont{Koester, Ismail, and
  Chu}}]{koester97}
\bibinfo{author}{\bibfnamefont{S.}~\bibnamefont{Koester}},
  \bibinfo{author}{\bibfnamefont{K.}~\bibnamefont{Ismail}}, \bibnamefont{and}
  \bibinfo{author}{\bibfnamefont{J.}~\bibnamefont{Chu}},
  \bibinfo{journal}{Semicond. Sci. Technol.} \textbf{\bibinfo{volume}{12}},
  \bibinfo{pages}{384} (\bibinfo{year}{1997}).

\bibitem[{\citenamefont{Lai et~al.}(2004)\citenamefont{Lai, Pan, Tsui, Lyon,
  M{\"u}hlberger, and Sch{\"a}ffler}}]{lai04}
\bibinfo{author}{\bibfnamefont{K.}~\bibnamefont{Lai}},
  \bibinfo{author}{\bibfnamefont{W.}~\bibnamefont{Pan}},
  \bibinfo{author}{\bibfnamefont{D.~C.} \bibnamefont{Tsui}},
  \bibinfo{author}{\bibfnamefont{S.}~\bibnamefont{Lyon}},
  \bibinfo{author}{\bibfnamefont{M.}~\bibnamefont{M{\"u}hlberger}},
  \bibnamefont{and}
  \bibinfo{author}{\bibfnamefont{F.}~\bibnamefont{Sch{\"a}ffler}},
  \bibinfo{journal}{Phys. Rev. Lett.} \textbf{\bibinfo{volume}{93}},
  \bibinfo{pages}{156805} (\bibinfo{year}{2004}).

\bibitem[{\citenamefont{Kharche et~al.}(2007)\citenamefont{Kharche, Prada,
  Boykin, and Klimeck}}]{kharche07}
\bibinfo{author}{\bibfnamefont{N.}~\bibnamefont{Kharche}},
  \bibinfo{author}{\bibfnamefont{M.}~\bibnamefont{Prada}},
  \bibinfo{author}{\bibfnamefont{T.~B.} \bibnamefont{Boykin}},
  \bibnamefont{and} \bibinfo{author}{\bibfnamefont{G.}~\bibnamefont{Klimeck}},
  \bibinfo{journal}{Appl. Phys. Lett.} \textbf{\bibinfo{volume}{90}},
  \bibinfo{pages}{092109} (\bibinfo{year}{2007}).

\bibitem[{\citenamefont{Goswami et~al.}(2007)\citenamefont{Goswami, Slinker,
  Friesen, McGuire, Truitt, Tahan, Klein, Chu, Mooney, der Weide
  et~al.}}]{goswami07}
\bibinfo{author}{\bibfnamefont{S.}~\bibnamefont{Goswami}},
  \bibinfo{author}{\bibfnamefont{K.}~\bibnamefont{Slinker}},
  \bibinfo{author}{\bibfnamefont{M.}~\bibnamefont{Friesen}},
  \bibinfo{author}{\bibfnamefont{L.}~\bibnamefont{McGuire}},
  \bibinfo{author}{\bibfnamefont{J.}~\bibnamefont{Truitt}},
  \bibinfo{author}{\bibfnamefont{C.}~\bibnamefont{Tahan}},
  \bibinfo{author}{\bibfnamefont{L.}~\bibnamefont{Klein}},
  \bibinfo{author}{\bibfnamefont{J.}~\bibnamefont{Chu}},
  \bibinfo{author}{\bibfnamefont{P.}~\bibnamefont{Mooney}},
  \bibinfo{author}{\bibfnamefont{D.~V.} \bibnamefont{der Weide}},
  \bibnamefont{et~al.}, \bibinfo{journal}{Nat. Phys.}
  \textbf{\bibinfo{volume}{3}}, \bibinfo{pages}{41} (\bibinfo{year}{2007}).

\bibitem[{\citenamefont{Takahashi et~al.}(1995)\citenamefont{Takahashi, Nagase,
  Namatsu, Kurihara, Iwdate, Nakajima, Horiguchi, Murase, and
  Tabe}}]{takahashi95}
\bibinfo{author}{\bibfnamefont{Y.}~\bibnamefont{Takahashi}},
  \bibinfo{author}{\bibfnamefont{M.}~\bibnamefont{Nagase}},
  \bibinfo{author}{\bibfnamefont{H.}~\bibnamefont{Namatsu}},
  \bibinfo{author}{\bibfnamefont{K.}~\bibnamefont{Kurihara}},
  \bibinfo{author}{\bibfnamefont{K.}~\bibnamefont{Iwdate}},
  \bibinfo{author}{\bibfnamefont{Y.}~\bibnamefont{Nakajima}},
  \bibinfo{author}{\bibfnamefont{S.}~\bibnamefont{Horiguchi}},
  \bibinfo{author}{\bibfnamefont{K.}~\bibnamefont{Murase}}, \bibnamefont{and}
  \bibinfo{author}{\bibfnamefont{M.}~\bibnamefont{Tabe}},
  \bibinfo{journal}{Electronics Letters} \textbf{\bibinfo{volume}{31}},
  \bibinfo{pages}{136} (\bibinfo{year}{1995}).

\bibitem[{\citenamefont{Zhuang et~al.}(1998)\citenamefont{Zhuang, Guo, and
  Chou}}]{zhuang98}
\bibinfo{author}{\bibfnamefont{L.}~\bibnamefont{Zhuang}},
  \bibinfo{author}{\bibfnamefont{L.}~\bibnamefont{Guo}}, \bibnamefont{and}
  \bibinfo{author}{\bibfnamefont{S.~Y.} \bibnamefont{Chou}},
  \bibinfo{journal}{\apl} \textbf{\bibinfo{volume}{72}}, \bibinfo{pages}{1205}
  (\bibinfo{year}{1998}).

\bibitem[{\citenamefont{Rokhinson et~al.}(2002)\citenamefont{Rokhinson, Tsui,
  Pfeiffer, and West}}]{rokhinson02}
\bibinfo{author}{\bibfnamefont{L.~P.} \bibnamefont{Rokhinson}},
  \bibinfo{author}{\bibfnamefont{D.~C.} \bibnamefont{Tsui}},
  \bibinfo{author}{\bibfnamefont{L.~N.} \bibnamefont{Pfeiffer}},
  \bibnamefont{and} \bibinfo{author}{\bibfnamefont{K.~W.} \bibnamefont{West}},
  \bibinfo{journal}{Superlattices Microstruct.} \textbf{\bibinfo{volume}{32}},
  \bibinfo{pages}{99} (\bibinfo{year}{2002}), \eprint{cond-mat/0303011}.

\bibitem[{\citenamefont{Bo et~al.}(2006)\citenamefont{Bo, Rokhinson, Yao, Tsui,
  and Sturm}}]{bo06}
\bibinfo{author}{\bibfnamefont{X.-Z.} \bibnamefont{Bo}},
  \bibinfo{author}{\bibfnamefont{L.}~\bibnamefont{Rokhinson}},
  \bibinfo{author}{\bibfnamefont{N.}~\bibnamefont{Yao}},
  \bibinfo{author}{\bibfnamefont{D.}~\bibnamefont{Tsui}}, \bibnamefont{and}
  \bibinfo{author}{\bibfnamefont{J.}~\bibnamefont{Sturm}}, \bibinfo{journal}{J.
  Appl. Phys.} \textbf{\bibinfo{volume}{100}}, \bibinfo{pages}{94317 }
  (\bibinfo{year}{2006}).

\bibitem[{\citenamefont{Rokhinson et~al.}(2000)\citenamefont{Rokhinson, Guo,
  Chou, and Tsui}}]{rokhinson00a}
\bibinfo{author}{\bibfnamefont{L.~P.} \bibnamefont{Rokhinson}},
  \bibinfo{author}{\bibfnamefont{L.~J.} \bibnamefont{Guo}},
  \bibinfo{author}{\bibfnamefont{S.~Y.} \bibnamefont{Chou}}, \bibnamefont{and}
  \bibinfo{author}{\bibfnamefont{D.~C.} \bibnamefont{Tsui}},
  \bibinfo{journal}{\apl} \textbf{\bibinfo{volume}{76}}, \bibinfo{pages}{1591}
  (\bibinfo{year}{2000}).

\bibitem[{\citenamefont{Klimeck et~al.}(2002)\citenamefont{Klimeck, Oyafuso,
  Boykin, Bowen, and von Allmen}}]{klimeck02}
\bibinfo{author}{\bibfnamefont{G.}~\bibnamefont{Klimeck}},
  \bibinfo{author}{\bibfnamefont{F.}~\bibnamefont{Oyafuso}},
  \bibinfo{author}{\bibfnamefont{T.}~\bibnamefont{Boykin}},
  \bibinfo{author}{\bibfnamefont{R.}~\bibnamefont{Bowen}}, \bibnamefont{and}
  \bibinfo{author}{\bibfnamefont{P.}~\bibnamefont{von Allmen}},
  \bibinfo{journal}{Comput. Model. Eng. Sci.} pp. \bibinfo{pages}{601 -- 42}
  (\bibinfo{year}{2002}).

\bibitem[{\citenamefont{Klimeck et~al.}(2007)\citenamefont{Klimeck, Ahmed, Bae,
  Kharche, Clark, Haley, Lee, Naumov, Ryu, Saied et~al.}}]{klimeck07}
\bibinfo{author}{\bibfnamefont{G.}~\bibnamefont{Klimeck}},
  \bibinfo{author}{\bibfnamefont{S.}~\bibnamefont{Ahmed}},
  \bibinfo{author}{\bibfnamefont{H.}~\bibnamefont{Bae}},
  \bibinfo{author}{\bibfnamefont{N.}~\bibnamefont{Kharche}},
  \bibinfo{author}{\bibfnamefont{S.}~\bibnamefont{Clark}},
  \bibinfo{author}{\bibfnamefont{B.}~\bibnamefont{Haley}},
  \bibinfo{author}{\bibfnamefont{S.}~\bibnamefont{Lee}},
  \bibinfo{author}{\bibfnamefont{M.}~\bibnamefont{Naumov}},
  \bibinfo{author}{\bibfnamefont{H.}~\bibnamefont{Ryu}},
  \bibinfo{author}{\bibfnamefont{F.}~\bibnamefont{Saied}},
  \bibnamefont{et~al.}, \bibinfo{journal}{IEEE Trans. Electron. Devices} pp.
  \bibinfo{pages}{2079 -- 89} (\bibinfo{year}{2007}).

\end{thebibliography}

\end{document}